# Controlling Swarms: A Programming Paradigm with Minimalistic Communication

**Joshua Cherian Varughese\*** · **Hannes Hornischer\*** · **Payam Zahadat** · **Ronald Thenius** · **Franz Wotawa** · **Thomas Schmickl**

**Abstract** Inspired by natural swarms, numerous control schemes enabling robotic swarms, mobile sensor networks and other multi-agent systems to exhibit various self-organized behaviors have been suggested. In this work, we present a Wave Oriented Swarm Programming Paradigm (WOSPP) enabling the control of swarms with minimalistic communication bandwidth in a simple manner, yet allowing the emergence of diverse complex behaviors and autonomy of the swarm. Communication in the proposed paradigm is based on "ping"-signals inspired by strategies for communication and self organization of slime mold (*dictyostelium discoideum*) and fireflies (*lampyridae*). Signals propagate as information-waves throughout the swarm. We show that even with 1-bit bandwidth communication between agents suffices for the design of a substantial set of behaviors in the domain of essential behaviors of a collective. Ultimately, the reader will be enabled to develop and design a control scheme for individual swarms.

## 1 Introduction

Nature teems with various kinds of life forms with varying individual capabilities. Many of these lifeforms have been found to depend not only on individual capabilities but especially on emergent group dynamics. Moving around as a swarm of individuals permits hunting together, foraging more efficiently, sharing food or

---

\* these authors contributed equally to this work; order of their listed names was chosen randomly

---

Joshua Cherian Varughese, Hannes Hornischer, Ronald Thenius and Thomas Schmickl
Institut für Biologie
Karl Franzens Universität Graz
E-mail: hannes.hornischer@uni-graz.at

Joshua Cherian Varughese and Franz Wotawa
Institut für Softwaretechnologie
Technische Universität Graz
E-mail: joshua.varughese@uni-graz.at

collaborative defense against predators to increase their collective probability of survival and reproduction [14]. Fish increases their individual probability of survival by moving in schools and collectively performing escape maneuvers when a predator is detected by one of the fish [5, 16]. The foraging efficiency of a hive increases dramatically when bees perform waggle dances [43, 35] to inform other bees about food sources in the vicinity. The synchronized blinking of fireflies [6, 7] and the aggregation of slime mold cells to form a slug [13, 36] are other examples of lifeforms relying on collective abilities for foraging and reproduction.

Various kinds of lifeforms with varying physical and cognitive complexity evolved to perform decentralized behaviors in order to ensure greater probability of survival. The part each individual has to play in such emergent behaviors is often simple, yet the result that emerges on a group level is resilient to failures of individuals or other disruptive events [23]. Numerous natural swarms following simple behaviors of its individuals are scalable and can therefore consist of hundreds, thousands or even millions of individual entities [20, 37].

The simplicity and decentralized nature of the individual's behaviors producing resilient collective phenomena have attracted interest from engineers seeking to inherit properties of simplicity, resilience, scalability in engineered systems. Several optimization algorithms relying on emergent behaviors have been proposed and successfully implemented[34, 19, 44]. "Particle swarm optimization" [19] is an optimization algorithm, inspired by flocks of birds, which is being used for multi objective optimization. BEECLUST [34] is an optima finding algorithm, inspired by a swarm of newly hatched bees, that can be used to find global optima. Numerous algorithms [32, 21, 17, 45] enabling simple agents to accomplish complex tasks such as source localization [17], task allocation [21, 45], collective mapping et cetera have been developed. In [32, 12], swarm intelligent algorithms are proposed that enable robots with very limited individual abilities to transport large objects in a collective manner. In [17], a distributed algorithm for localizing the source of an odor in an environment is proposed and tested on a swarm of robots. Apart from enabling a group of simple robots to perform complex tasks, various algorithms and methods have been suggested to control a swarm of robots of varying sizes to perform specific actions and tasks in a coordinated manner such as arranging themselves in a particular shape or responding collectively to external cues or stimuli [9, 22, 33]. In this paper a paradigm for controlling a group of agents is introduced and presented, using communication based on waves of single bit signals, or "pings" propagating through the swarm. The single bit pings propagating through the swarm is analogous to scroll waves of cAMP signals propagating through a swarm of *dictyostelium discoideum* while the periodic initiation of ping signals is inspired by *lampyridae*. WOSPP enables a swarm of agents with directional communication and locomotive capabilities to synchronize, elect a leader, estimate the number of members in the swarm, localize the center of the swarm, aggregate etc. In addition, a meta control scheme that enables a user to combine individual behaviors to produce more complex swarm behaviors is presented. In order to put this work into context of the current state of the art, related algorithms and approaches will be discussed briefly in Section 2.

Generally, in contrast to existing approaches for swarm control, the paradigm presented in this work allows a swarm to inherit rich self organized collective behaviors. Instead of complex messages or encoded signals solely one-bit communication suffices for the presented behaviors and ultimately allows the design of both top-down control interfaces and autonomous swarms. The basic communication behavior of the paradigm presented in this paper has already been used for producing gradient taxis[42]. Its basic concept as well as several behaviors have been explored in [39].

In this paper, a substantial set of behaviors in the domain of essential behaviors of a collective is presented and an extensive literature review is conducted in order to place the presented paradigm into perspective of existing work. For all presented behaviors the detailed structure and its design is described giving the reader an intuitive understanding of how to develop and design basic behavioral building blocks, which will be referred to as "primitives".

Since each individual application of swarm control requires a specific set of behavioral abilities, depending on the swarm members' abilities, their environmental conditions and their tasks the primitives have to be constructed or adjusted in a fitting manner. Primitives serve as a basis for a meta control scheme and can be combined in various ways in order to produce complex collective behaviors as shown in Figure 1. Ultimately, the reader will be enabled to design primitives and thus construct a meta control scheme for a swarm.

In Section 2, a literature review of relevant research is presented. Subsequently, in Section 3 the bioinspiration for the communication mechanism from slime mold and fireflies is presented. In Section 4, the communication mechanism of the paradigm presented in this paper is introduced as well as the fundamental concept of the paradigm. In Section 5 a set of primitives is introduced, classified into three categories: internal organization, swarm awareness and locomotion. In Section 6 possible methods for combining primitives and resulting complex behaviors are presented[1].

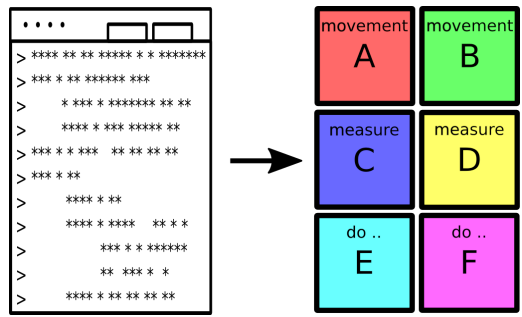

**Fig. 1** Schematic illustration of a meta control scheme. Every agent in WOSPP follows a fixed basic operational structure. Behavioral changes are only introduced through minor changes in two designated "codeBlocks", as presented later. This structure allows the coherent design of fundamental behaviors or "primitives" and thus the development of a meta control interface for controlling the swarm.

---

[1] All primitives presented in this paper have been simulated using Python 3.6.3

## 2 Related Research

In the field of controlling groups of entities, there is already a variety of approaches and previous work with diverse foci.

In [22], a classical approach to controlling a swarm is employed where users can interact with and control a swarm of certain robots, referred to as "zooids", using gestures. The authors achieved a responsive swarm using an external projector for tracking the robots position, assigning a goal position for each individual robot and then utilizing classical motion control strategies like Proportional-Integral-Derivative (PID) control to move the agents precisely from the start position to the goal position. Where as the programmer has control of the precise movements of the swarms members, the presence of a higher organizational entity is inherently necessary and thus depicts a classical example of top down control. There is a substantial amount of work related to controlling groups of entities. In the pioneering work of Craig Reynolds, he introduced self propelled particles known as boids [30] which exhibit self organized flocking and collision avoidance. Reynold's boids are able to mimic a flock of birds whose individuals follow simple behavioral rules. Due to its simplicity and decentralized structure it is applicable to large swarms. Its focus is on the generation of realistic behaviors as found in natural flocks and hence limited in its versatility.

In [26] a concept of self propelled particles with internal oscillators, or "swarmallators" is introduced. Attractive and repulsive forces are then used in relation to the relative phase shift of the oscillators for generating a range of collective behaviors. The internal processes' and states of the swarms' entities substantially influence and determine the interplay between individuals producing a small set of collective phenomena.

In [24], the authors introduce an algorithm for self assembly of identical agents on a surface into a predetermined global shape. Multiple gradients are developed by propagating messages starting from the agents at the edges in order to develop a relative positioning system among the agents. Subsequently, various shapes can be generated by manipulating the behavior of agents with particular gradient values. More complex shapes are achieved by repeating the process of generating gradients and folding along the specific areas of interest. A variation of the aforementioned idea was used to assemble various shapes in a self organized manner in a thousand robot swarm [33]. In [1], programmable self assembly and other similar research done by various researchers were unified as amorphous computing.

Among existing work, the paradigm presented in this paper exhibits most parallel characteristics with the approaches presented under amorphous computing, however significantly differs in several points. Instead of multi-bit signals encoding information, as used for communication in amorphous computing, the presented paradigm already produces rich behavioral diversity with single bit communication. Another key difference between is that the presented paradigm refrains from using "seed" agents[25] or global knowledge regarding edges and vertices in case of programmable self assembly for origami generation. Instead agents initiate communication with the rest of the swarm decentralized and randomly. In amorphous computing, languages such as Origami Shape Language (OSL) [24], Growing Point Language (GPL) [10] are utilized in order to enable a user to program a swarm. In this paper, the proposed paradigm will be used to program a swarm to perform collective behaviors and additionally, a meta control scheme can be designed for the swarm to perform these behaviors autonomously.

## 3 Bioinspiration

The programming paradigm presented in this paper takes inspiration from the communication mechanisms used by slime mold and fireflies. In the following the relevant aspects of these lifeforms are briefly discussed.

### 3.1 Slime mold

Slime mold (*dictyostelium discoideum*), is a free living diploid life form which takes advantage of swarming behavior to survive challenging environments. Slime mold has a widely varying cooperation with other cells during its life cycle depending on the environment. When there are ample food sources, cells grow and divide individually without cooperation. In case of scarcity of food or other threats, significant cooperation between the cells begin. During its cooperative phases, the individual cells aggregate to form a multicellular organism. And the collective begins to act similar to a single organism. The paradigm presented in this paper will chiefly consider the signaling behavior of slime mold during the aggregation of slime mold cells. For aggregating, some cells (centers) release Cyclic Adenosine Monophosphate (cAMP) into the environment to recruit surrounding cells to join the aggregate. This signal induces a short-lived chemical concentration spike around the recruiting cells [36]. Other slime mold cells that perceive the chemical signal will produce the emit the chemical themselves to relay the signal. Since all slime mold cells relay the chemical signals they receive, the the original signal produced by the recruiting center rapidly travels through the swarm. Additionally, each cell needs around 12-15 seconds [2] in between two cAMP signals. During this insensitive period, individual cells are insensitive to any cAMP pulses. This intermediate insensitive time acts as a "refractory" period that prevents any echoing between two amoeba cells. The signal relaying mechanism described above forms the basis for spatio-temporal patterns known as scroll waves. Since the origin of the waves are at the recruiting cells, the amoeba can move towards the incoming signal to reach the recruiting center [36].

### 3.2 Fireflies

Fireflies (*lampyridae*) are a family of insects that are capable of producing bioluminescence to attract a mate or a prey [6]. Bio-luminescence of various families of fireflies has been a subject of elaborate study in the past [6]. Apart from being able to blink, fireflies are known to behave in cooperation with other fireflies in order to attract mates or prey [6]. Such synchronicity is a result of a simple mechanism by which initially the individual fireflies blink periodically. When a firefly perceives a blink in its surrounding, it blinks again and then resets its own blinking frequency to match the received blink [8]. This is analogous to a phase coupled oscillator which adjusts its phase to match it to that of the faster one in the vicinity. This trait emerges into a quasi synchronized blinking pattern while the frequency of blinking will be influenced by the fastest blinking insect.

## 4 WOSPP - Wave Oriented Swarm Programming Paradigm

The wave oriented swarm programming paradigm WOSPP is strongly inspired by the two aforementioned organisms slime mold and fireflies.

In particular, communication within the paradigm is based on waves propagating through a swarm. Every agent within a swarm has the ability of sending and receiving information signals, which we will refer to as "pings", to nearby agents. Connected to this, all agents can enter three different states, parallel to the behavior of slime mold: An inactive state in which agents are receptive to incoming signals (responsive to cAMP or "pings"), an active state where they send or relay a signal (release cAMP or send "ping") and optionally perform an action, which is followed by a refractory state where agents are temporarily insensitive to incoming signals. This is schematically shown in Figure 2(a).

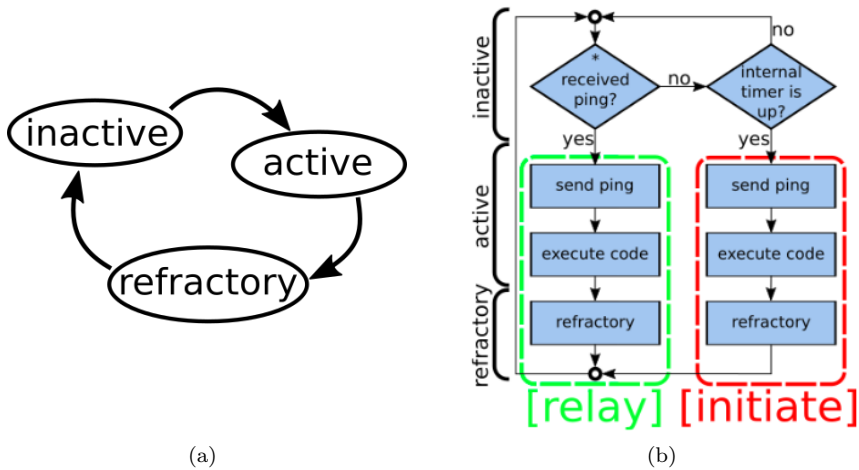

(a)              (b)

**Fig. 2** (a): Three states of agents in WOSPP: From inactive state, through external trigger (e.g. incoming ping) or internal trigger (timer) the agent transitions into the active state where it sends a ping and optionally performs an action. Afterwards, it enters the refractory state, being insensitive to incoming signal until transitioning into the inactive state again. (b):The conceptual operating structure of an individual agent. If an agent in the inactive state receives a ping, it relays the signal by entering the active state and sending a ping itself. Specific to the primitive a certain code will be executed and the refractory state is entered. If an agent's internal time is up it will initiate a ping following the same structure, however executing a different code (specific to the primitive) in the active state.

This operational structure results in a wave like propagation of signals throughout the swarm. Agents in the inactive state get triggered to relay a signal, while the refractory state prevents the system from continuously signaling and thus flooding the system. In Figure 4 the propagation of waves is shown for a swarm of agents, each agent represented by a dot with the color denoting their state. The inactive state is denoted in black, the active state in red and the refractory state in green.

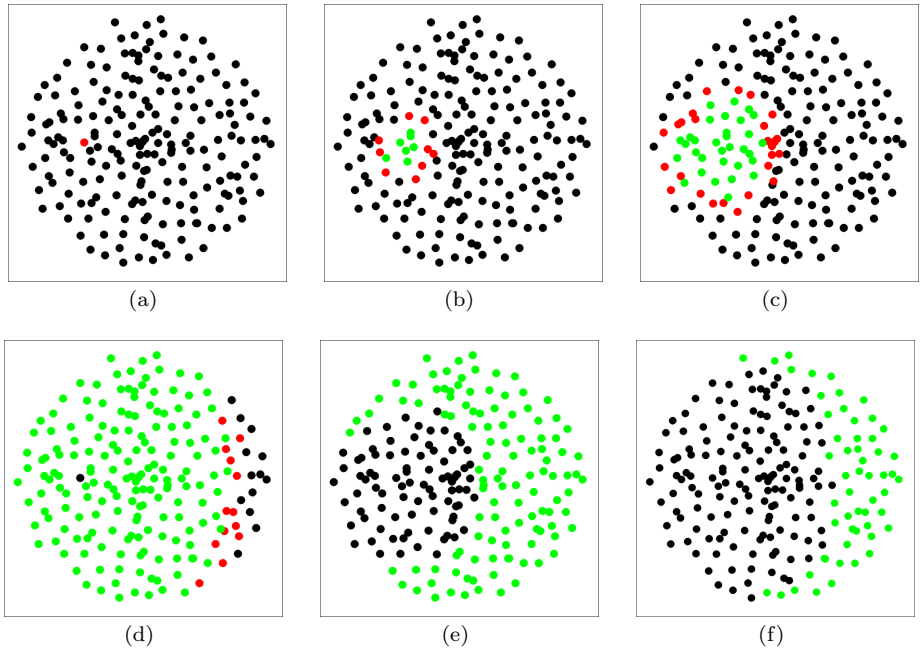

**Fig. 3** Illustration of wave based communication. In (a) almost all agents are in the inactive state, shown in black, except one agent which broadcasts a ping, i.e. enters the active state, shown in red. It afterwards transitions into the refractory state, shown in green. Neighboring agents receive the signal and themselves transition into the active state as shown in (b) and (c). The ping signal spreads in a wave like manner. In (d) the initiating agent transitions from the refractory state into the inactive state again. Due to a fixed duration of the refractory state, the transition to the inactive state as well spreads in a wave like manner, shown in (e) and (f). Times [s]: (a) 0, (b) 2, (c) 4, (d) 11, (e) 16, (f) 19. Parameters (as defined in more detail at the end of Section 4): number of agents $N = 80$, physical size of the swarm in units perception range $R_s = 5\,r$, refractory time in units timesteps $t_{ref} = 5\,s$.

The ability of fireflies to adjust and reset their individual frequency of "blinking" is ground to a concept used in this paradigm: internal timers. Every agent has an internal timer which, when running out, will trigger the agent to enter the "active" state where it broadcasts a ping, thus initiating a ping wave through the swarm. For most primitives, this timer is reset right after running out, causing an agent to repeatedly count down and, subsequently, ping.

This communication structure and the internal timers constitute the basic and fixed structure of an agent in WOSPP and is shown as pseudo code in Algorithm 1. This basic structure suffices for the behavior shown in Figure 4 and is fixed for all agents. However, as will be presented later in this work, complex behavior can be induced by adding simple command in the codeBlocks. When agents perform simple actions when relaying or initiating pings, complex behavior can emerge. This structure is conceptually shown in a flowchart in Figure 2(b). The three states are marked, as well as two separate operational chains, one for relaying, one for initiating pings. Both incorporate an optional field for executing the respective codeBlocks.

For the behaviors, or primitives, presented in this work, agents are not only able to send and receive pings, but also optionally have a heading and sense of directionality. Agents are enabled to determine the direction from which they receive a ping. Furthermore, for some tasks agents have the ability to move in direction of their heading. Primitives in which the agents move requires agents to move sufficiently slow in order to avoid single agents being left behind or splitting the swarm up into smaller groups. The agents are considered as point particles and thus collision detection is ignored within this work since it is highly specific for each individual swarm and its environmental conditions. Aside from primitives shown in this paper a large variety of behaviors can be developed, incorporating abilities and environmental sensors of agents not discussed here. This work gives a fundamental introduction into the concept for enabling a modular design and development of primitives for swarm control tailored to specific conditions and tasks of a swarm.

Regarding communication bandwidth, for the individual primitives presented in this work exclusively single-bit communication is considered in order to show that even for the minimum communication case remarkable behaviors can be produced. The use of multi-bit communication instead of pings would further increase the possibilities for swarm control. Multi-bit communication is discussed in Section 6 and in the discussion.

For the simulations presented in this work only swarms with agents spatially distributed in two dimensions in an approximately circular shape were used. This is chosen as a general shape without loss of generality, it does not limit or affect functionality of WOSPP regarding shape or spatial dimension of swarms.

In the following parameters and quantities used in this work are introduced and defined.

- In the numerical simulations presented here, time is measured in timesteps $s$. An agent receiving a ping will be activated and itself send a ping one timestep $t_{activate} = 1\,s$ later.
- Every agent has an internal timer $t_p$ which usually periodically resets to a maximum $t_p^{max}$. However, for some primitives however the timer can be reset to a random number between $t_p \in (0, t_p^{max}]$.
- The number of agents constituting a swarm is defined as $N$. The minimum $N$ necessary for the presented primitives to function is $N = 2$. The maximum can theoretically be arbitrarily large, as further elaborated in the discussion.
- The refractory time $t_{ref}$ denotes the time an agent remains in the refractory state, i.e. insensitive to incoming pings. For the presented simulations $t_{ref}$ is set to be larger than the time a signal would take to propagate at the edge of the swarm in a circular manner. This is for preventing a ping wave from continuously propagating through the swarm.
- For distances the basic unit is perception range of agents $r$, the distance up to which an agent will perceive the pinging of a nearby agent.
- For primitives including locomotion, agents take discrete spatial steps within a timestep, where the length of their step $d$ is set to one-sixth of a perception range $d = r/6$.

– The physical size of the swarm is defined as $R_s$ and will be given in units perception range $r$.

**Data:** Paradigm parameters
**Result:** -
state $\leftarrow$ *inactive*;
timer($t_p$) $\leftarrow$ random integer $\in (0,\ t_p^{max}]$;
**while** *primitive* **do**
    decrement timer($t_p$);
    **if** *agent in refractory state* **then**
        wait for refractory_time;
        **if** *refractory_time is over* **then**
            state $\leftarrow$ inactive
        **end**
    **end**
    **if** *agent in active state* **then**
        broadcast ping;
        state $\leftarrow$ refractory
    **end**
    **if** *agent in inactive state* **then**
        listen for incoming pings;
        **if** *ping received* **then**
            state $\leftarrow$ active;
            execute Relay-CodeBlock;
        **end**
    **end**
    **if** *timer($t_p$) $\leq$ 0* **then**
        state $\leftarrow$ active;
        execute Initiate-CodeBlock;
    **end**
**end**
**Function** *Initiate-CodeBlock*
    -
**Function** *Relay-CodeBlock*
    -

**Algorithm 1:** Basic pseudo code for every individual agent within WOSPP. This structure is fixed, behavioral changes are only introduced through adding commands to the initiate- and relay-codeBlocks, here highlighted. The timer $t_p$ is initially set to a random value with upper limit $t_p^{max}$.

## 5 Primitives

In this section a set of primitives is presented where small changes in the code-Blocks produce large scale complex behavior. For every primitive plots of results for intuitively visualizing the behavior are presented, as well as the code-block. The presented primitives are divided into three categories:

1. **Internal organization** is about self organization of the swarm on an internal level of each agent, including the primitives "leader election", "synchronization" and "localize object".
2. **Swarm awareness** includes the individuals awareness about properties of the swarm or properties of itself within the swarm. Presented are the primitives "localize swarm center", "estimating number of swarm members" and "estimate individual position within swarm".
3. The category **locomotion** is about physically self organizing or restructuring the swarm, including the primitives "aggregation", "moving collectively" and "gas expansion".

In the following, if not stated otherwise, every swarm is initially randomly distributed within a circular area of radius $R_s$ such that every agent is connected to at least one neighbor.

### 5.1 Internal organization: Leader election

For various tasks it can be beneficial or even necessary for a swarm to have a certain agent "leading" a swarm. Having a certain agent assigned as a special entity brings the risk of having this agent removed from a swarm and thus disabling the entire swarm. Instead the swarm can collectively elect a leader thus eliminating such risks. For deciding for a leader, initially all agents consider themselves potential leaders, shown in Figure 4(a) in black. An agent pinging is illustrated in red and an agent not considering itself a leader anymore is green. Every agent sets its timer to a random number within $t_p \in (0, t_p^{max}]$. As soon as an agent receives a ping before its own internal timer ended it will not consider itself a candidate anymore and also deactivate its internal timer, i.e. not initiate pinging. After an agent initiated a ping it will again randomly choose a time $t_p$ for initiating pinging another time. This is shown as pseudo code in Algorithm 2.

**Data:** Paradigm parameters
**Result:** Leader election
.
.
**Function** *Initiate-CodeBlock*
| candidate ← true;
| timer($t_p$) ← random integer ∈ $(0, t_p^{max}]$;
**Function** *Relay-CodeBlock*
| deactivate internal timer;
| candidate ← false;

**Algorithm 2:** Code block for primitive "Leader election"

Figure 4(b) and (c) show agents initiating ping waves and immediately outrivaling their surrounding agents. The refractory mode prevents two initiating agents from outrivaling each other, however more than one can be left as potential leaders, as shown in Figure 4(d). Since every remaining candidate again chooses a random time to ping, after few "negotiation cycles" a single candidate, which then can be considered the leader, will remain, as shown in Figure 4 (e) and (f). Alternative approaches for leader election in groups e.g. based on voting [18] as well as decentralized probabilistic methods [4] can be found in the literature.

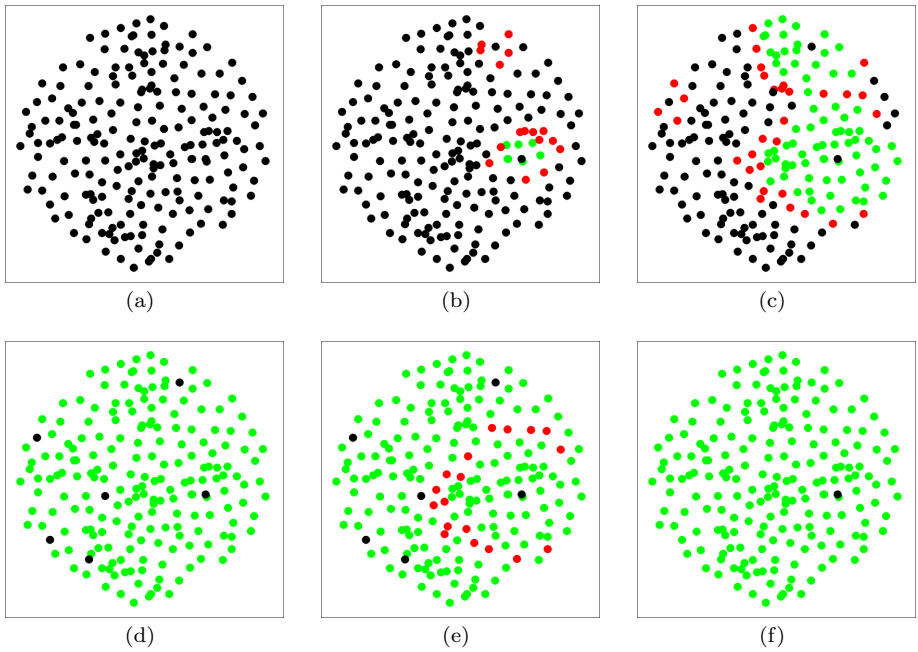

**Fig. 4** Leader election in a swarm. Candidates are shown in black, pinging agents in red and agents not considering themselves candidates in green. Initially all agents consider themselves candidates or potential leaders (a). After receiving and relaying a ping an agent will not consider itself a candidate anymore. Agents initiating pinging thus outrival agents around them. In (b)-(d) it is illustrated how several agents initiate pinging and not outrival each other (due to refractory time). Since only candidates will initiate pinging, the remaining candidates then repeat the process, indicated in (e) until only a single candidate remains, as shown in (f). Times [s]: (a) 0, (b) 5, (c) 8, (d) 30, (e) 41, (f) 189. Parameters: $N = 200$, $R_s = 6\,r$, $t_{ref} = 10\,s$, $t_p^{max} = 100\,s$.

## 5.2 Internal organization: Synchronization

For a swarm being able to perform coordinated actions it often is essential to synchronize regarding their respective tasks. This primitive allows the swarm members to synchronize regarding e.g. the sending of pings, allowing to perform actions quasi-simultaneously.

Every agent sets its internal counter to a random value between $t_p \in (0, t_p^{max}]$. If an agent receives a ping, it resets its internal counter to $t_p^{max}$. This is shown as pseudo code in Algorithm 3. As a result, the first agent sending a ping (which is then being relayed and propagates through the system wave-like) resets the timers of all relaying agents to the maximum $t_p^{max}$. Hence, the entire swarm will ping quasi-simultaneously within a time period smaller or equal to the duration of a ping propagating from one end of the swarm to the other. In Figure 5(a) the synchronization process for a swarm of $N = 15$ agents is shown via an order parameter $\Delta\phi_{max}$ which decreases with increasing synchronicity within the swarm. $\Delta\phi_{max}$ is calculated by determining the smallest phase interval containing the timers of all agents and then taking the maximum phase difference of all timer pairs. At $t = 30\,s$ the onset of synchronization is indicated with a gray vertical line. In Figure 5(b) the corresponding internal timers of all agents are shown, incrementally decreasing with time. Every line of points represents the internal timer of one agent. At $t = 30\,s$ an agent initiates pinging and thus resets the timers of all other agents such that at $t = 35\,s$ all agents reset and thus synchronized. Alternative approaches to synchronization in swarms employs pulse coupled oscillators inspired by fireflies [27, 28].

> **Data:** Paradigm parameters
> **Result:** Synchronized Swarm
> .
> .
> .
> **Function** *Initiate-CodeBlock*
> | timer($t_p$)← $t_p^{max}$;
> **Function** *Relay-CodeBlock*
> | timer($t_p$)← $t_p^{max}$;

**Algorithm 3:** Code block for primitive "Synchronization". The entire pseudo code for each agent is shown. The highlighted part is the pseudo code which differs for every primitive. For all other primitives presented here only the latter part will be shown.

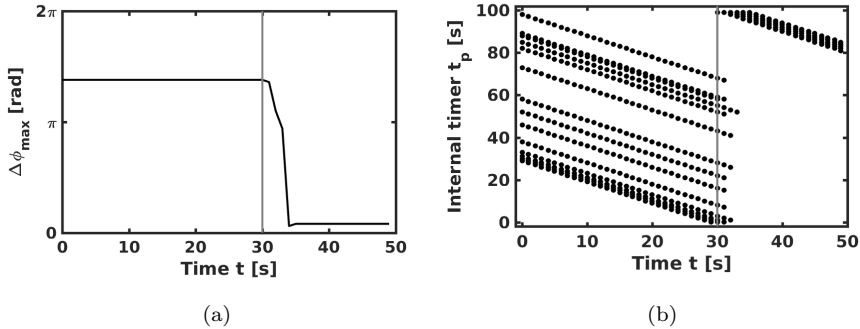

(a)                              (b)

**Fig. 5** (a): Onset of synchronized internal timers at $t \approx 30$ of a swarm of $N = 15$ agents, indicated with a gray vertical line. The order parameter $\Delta\phi_{max}$ is plotted against simulation time $t$. $\Delta\phi_{max}$ is calculated by determining the smallest phase interval containing the timers of all agents and then taking the maximum phase difference between two timers. After fully synchronizing at $t \approx 36\,s$ the maximum phase difference decreased from $\Delta\phi_{max} \approx 4.4\,$rad to $\Delta\phi_{max} \approx 0.25\,$rad which corresponds to a time interval of $\Delta t \approx 4\,s$. This interval can be identified in (b), where the internal timers of the agents versus simulation time is shown. Every line of data points corresponds to the internal timer of a single agent, which incrementally counts down. All timers decrease until at $t = 30\,s$ an agent's timer reaches $t_p = 0$ and thus initiates pinging. This marks the onset of the synchronization process and is marked with a gray vertical line. All agents relaying the ping then reset their timers. The reset signal propagates through the swarm and at $t = 35\,s$ all agents reset and collectively count down in a quasi-synchronous manner, that is, within a time interval of $\Delta t = 4\,s$. Parameters: $N = 15$, $R_s = 1.67\,r$, $t_{ref} = 20\,s$, $t_p^{max} = 100\,s$.

### 5.3 Internal organization: Localize object

For distributing information about spatial structure of the surrounding, a swarm needs to be able to communicate the location of nearby objects or events among its members. This primitive enables a swarm to collectively localize the direction of a direct path towards an object which one or few members of the swarm detect. Each agent refrains from initiating pinging unless it itself detects an object. Every agent receiving a ping, records the direction of the incoming ping. An estimate of the direction towards the object is then obtained by taking a running average of the directions of incoming pings. The pseudo code is shown in Algorithm 4.

Figure 6 shows the agents estimate of the rough location of the object as arrow, placed at the position of the agent within the swarm. The red square represents an object which can only be detected by agents in its vicinity. Figure 6 (a) shows the initial (random) orientation of the agents. With increasing number of perceived pings, the estimate of direction towards the object increases in accuracy until agents accurately point towards the position of the object as shown in Figure 6 (b).

**Data:** Paradigm parameters
**Result:** Agent knows rough direction of an object
.
.

**Function** *Initiate-CodeBlock*
> timer($t_p$) ← random integer ∈ $(0, t_p^{max}]$;
> **if** *no object detected* **then**
> > state ← inactive;
>
> **end**

**Function** *Relay-CodeBlock*
> record ping direction;
> current estimate ← average ping directions

**Algorithm 4:** Code block for primitive "Localize object"

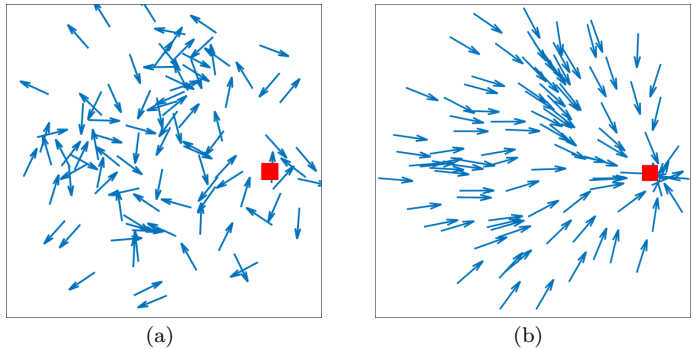

(a)                                              (b)

**Fig. 6** The arrows representing the agents estimation of the location of the object, the beginning of the arrow denotes the position af an agent. (a): The estimate is initialized to random direction at the start of the simulation. (b): The converged estimation of the location of the object after one $t_p^{max}$. All agents now point towards the object. Parameters: $N = 100$, $R_s = 3.34\,r$, $t_{ref} = 5\,s$, $t_p^{max} = 2500\,s$.

## 5.4 Swarm awareness: Localize swarm center

For a swarm being able to execute spatially coordinated actions the knowledge of the individual about the location of the center of the swarm can be of great advantage. This primitive enables each swarm member to identify the direction from where most signals originate from, which will be referred to as Average Origin of Pings, or AOP. For a swarm of the presented type (circular, approximately homogeneously distributed, agents have several communication neighbors) this direction coincides with the direction towards the physical center of the swarm. Each agent sets its internal counter $t_p$ to a random value between $t_p \in (0, t_p^{max}]$, as soon as a counter reaches $t_p = 0$ an agent will send a ping. When an agent receives a ping it stores the direction of the incoming ping and averages over all stored directions. This is shown as pseudo code in Algorithm 5.

**Data:** Paradigm parameters
**Result:** Agent knows rough direction of swarm center
.
.
**Function** *Initiate-CodeBlock*
 | center estimate ← mean of previous estimates;
 | timer($t_p$) ← random integer ∈ $(0, t_p^{max}]$;
**Function** *Relay-CodeBlock*
 | record ping direction;
 | current estimate ← average ping directions
 **Algorithm 5:** Code block for primitive "Localize swarm center"

After one cycle every agent once initiated a ping wave, unless a ping was initiated while all surrounding agents were in refractory state and thus the ping not relayed. Figure 7 shows a swarm in its initial state and after it equilibrated where every agent's orientation is denoted by an arrow at the position of the agent in the swarm. Initially the heading is random. After equilibrating, the agents on the outside accurately point towards the center of the swarm.

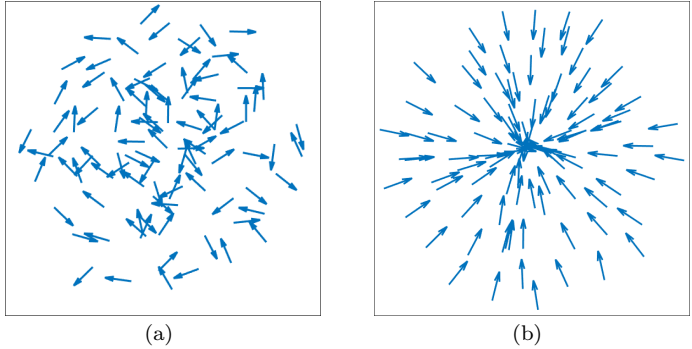

(a)           (b)

**Fig. 7** Agents estimation of the direction towards the center of the swarm. The beginning of an arrow denotes the position of an agent. (a) shows the initial estimates of each agent as arrow at its position in the swarm. (b) shows the converged estimates after $t = 4\,t_p^{max}$. Parameters: $N = 100$, $R_s = 3.34\,r$, $t_{ref} = 5\,s$, $t_p^{max} = 2500\,s$

5.5 Swarm awareness: Estimating number of swarm members

For some tasks a swarm may needs to be constituted of a certain number of agents in order to effectively operate. Or a swarm may needs check if the number of its members substantially changed, due to loss of members or merging with another swarm. This problem can be solved without an external observer as the swarm can estimate its number of swarm members autonomously. Each agent sets its internal counter $t_p$ to a random value between $t_p \in (0, t_p^{max}]$. Whenever an internal timer is up, an agent will initiate pinging and randomly reset its timer to $t_p \in (0, t_p^{max}]$. Each time an agent receives a ping it relays the signal and increments a counter $N_{count}$.

Furthermore, every time an agent initiates pinging, i.e. one internal cycle has passed, it will store its counter $N_{count}$ as its estimate of the number of swarm members for the past cycle. The average over those estimates will be the agent's opinion of the number of members in the swarm $N_{est}$. This is shown as pseudo code in Algorithm 6.

**Data:** Paradigm parameters
**Result:** Agent knows approximate number of members in the swarm
.
.
**Function** *Initiate-CodeBlock*
    | estimate($N_{est}$) ← mean of previous swarm size counters;
    | counter($N_{count}$) ← 0;
    | timer($t_p$) ← random integer ∈ $(0, t_p^{max}]$;
**Function** *Relay-CodeBlock*
    | increment counter $N_{count}$;
**Algorithm 6:** Code block for primitive "Estimating number of swarm members"

In Figure 8 the estimate $N_{est}$ averaged over all members of a swarm versus simulation time is shown. The estimate quickly increases before slowly converging to $N_{est} \approx 34$. The error bars represent the standard deviation thus indicating that the estimates of all agents are closely distributed around the mean. For the data in Figure 8, a swarm of 50 agents was building estimates over a time of $20 \cdot t_p^{max} = 20 \cdot 1000\,\mathrm{s}$, so $2 \cdot 10^4$ timesteps. The estimate converges to a value significantly lower than the actual number of swarm members however, for the same swarm the estimate consistently converges to the same (lower) estimate.

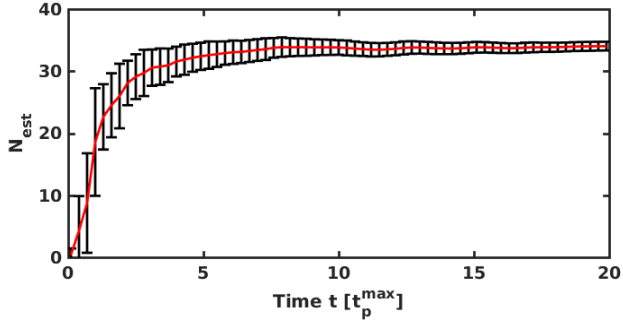

**Fig. 8** Estimated number of swarm members averaged over all agents in the swarm versus time. The error bars represent the standard deviation. The estimate steeply increases from $N_{est} = 0$ to $N_{est} = 30$ at $t = 3\,t_p^{max}$ before it gradually converges to its final estimate of $N_{est} \approx 34$. Parameters: $N = 50$, $R_s = 3.34\,r$, $t_{ref} = 5\,s$, $t_p^{max} = 1000\,s$.

Figure 9 shows the percentage deviation $N_{err}$ from the actual number of swarm members $N$ depending on the maximum possible cycle length $t_p^{max}$. For every data point the simulation was run for $25\,t_p^{max}$, sufficiently long for the estimate to converge.

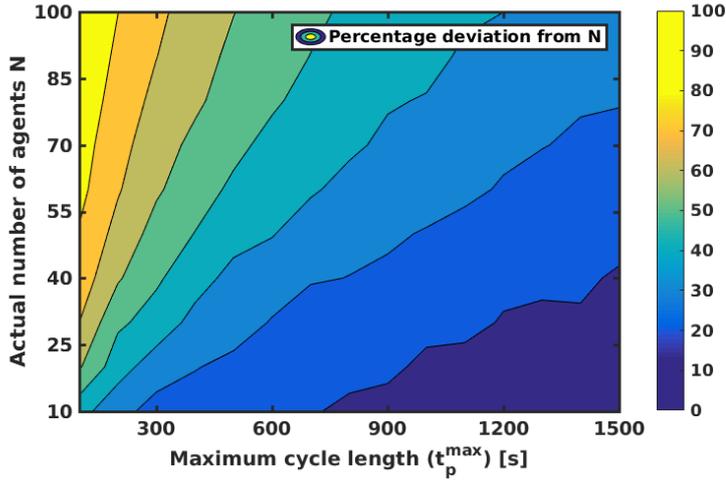

**Fig. 9** Percentage deviation of estimated number of agents in the swarm versus maximum cycle length and the actual number of agents in the swarm. Agents consistently underestimate the number of members of the swarm. The deviation decreases for decreasing $N$ and increasing $t_p^{max}$. Parameters used: $N \in \{10, 20, .., 100\}$, $R_s = 1\,r$, $t_{ref} = 5\,s$, $t_p^{max} \in 100, 200, .., 1500\,s$.

With increasing $t_p^{max}$ the deviation from the actual number of members of the swarm decreases. Knowing the order of magnitude of $N$ of a swarm, the $t_p^{max}$ can be chosen sufficiently large such that the deviation sufficiently small. For instance, considering a swarm of a maximum of 30 agents of the presented kind, a maximum cycle length of $t_p^{max} > 1500$ would ensure a deviation of around $N_{dev} = 10\%$.

However, a general corrective function for determining the approximate systematic deviation from the actual number of agents in the swarm can be developed, though this will not be further discussed in this work.

## 5.6 Swarm awareness: Estimate individual position within swarm

Some actions require the agents in a swarm to determine their relative position within the swarm. Considering for example a swarm protecting itself from environmental hazards, requiring only the agents at the edge or outer shell of the swarm to take measures, it is necessary for each individual agent to learn about their approximate relative position. For this, each agent sets its internal timer $t_p$ to a random value between $t_p \in (0, t_p^{max}]$. As previously explained, this will result in agents randomly pinging at random time slots and each agent relays the received pings to nearby agents. The agents then bin each of the pings received into four directions of $\alpha = 90°$ each. If there is at least one empty bin with no pings received, then the agents perceives itself as being on the periphery of the swarm. Pseudo code is shown in Algorithm 7.

---

**Data:** Paradigm parameters
**Result:** Agent knows if it is at the periphery
.
.
**Function** *Initiate-CodeBlock*
    **if** *Is at least one bin empty?* **then**
        | periphery ← true;
    **else**
        | periphery ← false;
    **end**
    timer($t_p$) ← random integer $\in (0, t_p^{max}]$;
**Function** *Relay-CodeBlock*
    record ping direction;
    bin incoming ping directions into bins of 90°;

**Algorithm 7:** Code block for primitive "Localize object"

---

Figure 10 shows the perception of agents regarding their position in the swarm. Initially, no agents perceive if they are at the periphery of the swarm, denoted through the black color of agents in Figure 10(a). As agents receive more pings from the surrounding agents, they are able to have more accurate estimate of its own position within the swarm as shown in Figure 10(b) where red colored agents perceive that they are at the periphery of the swarm.

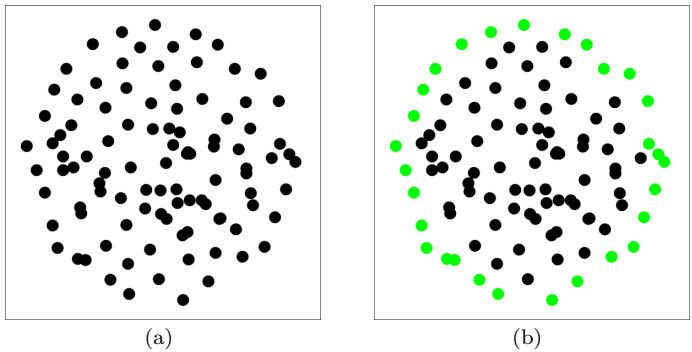

**Fig. 10** Figures show the agents perception of their location within the swarm. Red colored agents perceive that they are on the periphery and the black colored agents perceive them not being on the periphery. (a), shows the initialization at the start of the simulation with all agents perceiving themselves as "not being at the periphery". (b) shows the converged perception of the agents after 10 $t_p^{max}$. Parameters: $N = 100$, $R_s = 4\,r$, $t_{ref} = 5\,s$, $t_p^{max} = 2500\,s$

Another approach, enabling agents to estimate their distance from the average origin of pings is to let agents collect the number of incoming pings per directional bin and then merge the four bins into two in a way, such that two bins of $\alpha = 180°$ arise, one with the minimum count of received pings, the other half with the maximum. With increasing number of members in a swarm the accuracy of this approach increases. A large variety of approaches for localization within a swarm can be found in the literature. Some of the established methods use Kalman Filtering [31], Monte Carlo Localization [15] or localization based on local information [11].

### 5.7 Locomotion: Aggregation

Considering a swarm of agents with the ability to move and spatially arrange itself, for regrouping it needs to be able to gather or aggregate. For that every agent randomly sets its internal counter $t_p$ to a random value between $t_p \in (0, t_p^{max}]$. An agent receiving a ping will, after relaying it, move a small distance towards the incoming ping. This way, gradually all agents move towards each other. Pseudo code is shown in Algorithm 8.

Figure 11 shows a swarm aggregating in such manner. From its initial state the swarm steadily moves towards its average origin of pings, causing it to aggregate at the center of the swarm. Figure 11(d) shows the aggregated state of the agents as well as each agents trajectory as blue line. This illustrates how agents tend to follow the paths of their fellow members of the swarm, producing a root-like trajectory structure.

For illustrating the aggregation process, Figure 12 shows in blue the average root mean square distance of all agents to the center, i.e. the average position of all agents at that time. In red the same quantity averaged over 20 independent simulations is shown.

**Data:** Paradigm parameters
**Result:** Aggregated swarm
.
.

**Function** *Initiate-CodeBlock*
   |  timer($t_p$) ← random integer ∈ $(0, t_p^{max}]$;

**Function** *Relay-CodeBlock*
   |  timer($t_p$) ← $t_p^{max}$;
   |  record ping direction;
   |  Calculate average of incoming pings;
   |  move towards incoming ping;

**Algorithm 8:** Code block for primitive "Aggregation"

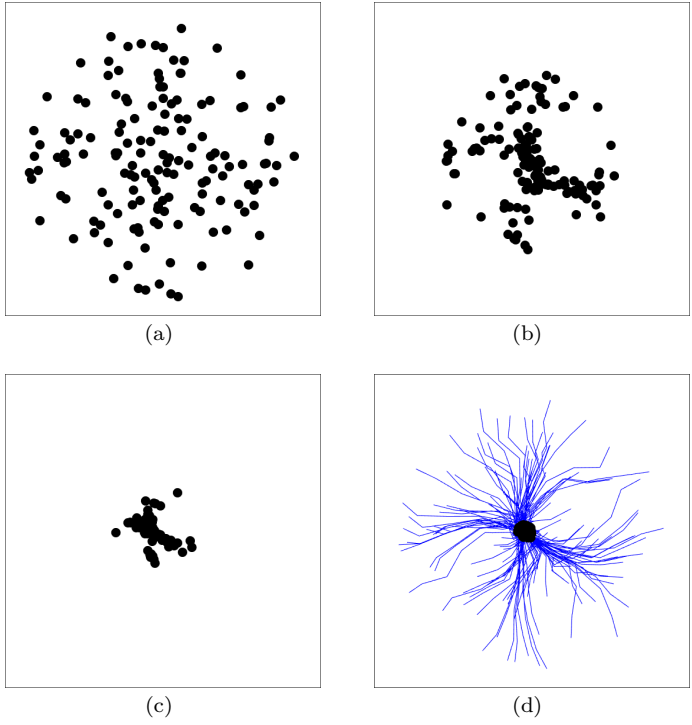

(a)                 (b)

(c)                 (d)

**Fig. 11** Aggregation of the swarm. Initial state of the swarm is shown in (a), in (b) and (c) it steadily aggregates. The final state and trajectories of each agent for the entire simulation as blue lines are shown in (d). Times $[t_p^{max}]$: (a) 0, (b) 0.7, (c) 1.4, (d) 2. Parameters: $N = 80$, $R_s = 2\,r$, $t_{ref} = 10\,s$, $t_p^{max} = 500\,s$.

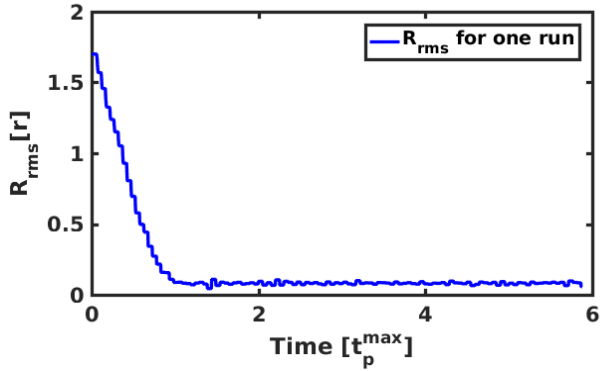

**Fig. 12** Average root mean square distance of all agents from the center of the swarm plotted against time. The blue graph shows the $R_{rms}$ of the swarm in the simulation shown in Figure 11. The $R_{rms}$ linearly decreases until $t_p^{max} \approx 1$ when the swarm almost fully aggregated. If looked at in close, the linear decrease occurs in quasi-discrete steps, corresponding to ping waves causing all agents to move towards each other quasi-simultaneously. Parameters: $N = 80$, $R_s = 2\,r$, $t_{ref} = 10\,s$, $t_p^{max} = 500\,s$.

Considering, a swarm needs to aggregate at a specific location, e.g. for measuring or exploring a certain area or object, the primitive can be changed such that only certain agents, which for example perceive stimuli such as the presence of an object, are able to initiate pinging. This is shown in Figure 13. The stimulus can also be an event or can be connected with a gradient. Considering agents with the ability to perceive e.g. light intensity, the agents will be able to aggregate at the brightest spot if every agent sets its internal counter to a value proportional to its perceived brightness. The agents at the brightest spots will statistically ping first. Furthermore, every agent receiving a ping will reset its counter thus allowing the agents at the brightest spot to hijack the swarm. This process executed repeatedly will result in a gradient taxis behavior as presented in [42]. It is worth noting that various approaches for aggregation in swarms have been developed and presented based on differing mechanisms and varying levels of complexity [40, 3, 34].

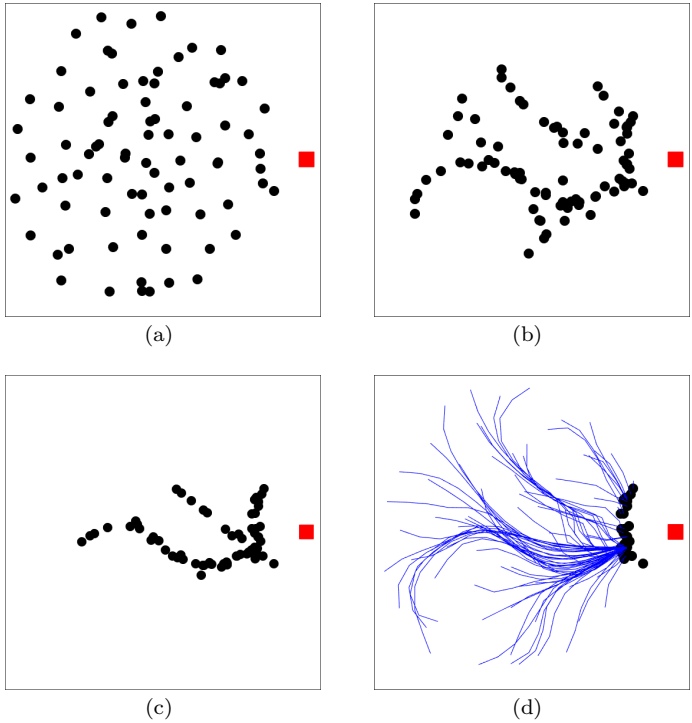

**Fig. 13** A swarm aggregating at an object, marked as red square on right hand side of the shown system. Initial state of the swarm is shown in (a). The swarm gradually aggregates at the object in (b) and (c) until every agent directly perceives the object in (d). Also shown are the trajectories for the entire simulation as blue lines. Times $[t_p^{max}]$: (a) 0, (b) 10, (c) 20, (d) 64. **make the object square and bigger to make it more different from agents.** Parameters: $N = 80$, $R_s = 3.34\,r$, $t_{ref} = 15\,s$, $t_p^{max} = 50\,s$.

## 5.8 Locomotion: Moving collectively

For the mobility of a swarm the ability to collectively move to a specific location can be crucial. For letting the entire swarm move towards a certain direction, a single agent serves as leader. Exclusively this leader initiates pings and gradually moves along a trajectory leading to the target location. All agents receiving pings will move towards the direction of it and thus follow the leader. The pseudo code is shown in Algorithm 9.

---

**Data:** Paradigm parameters
**Result:** Swarm follows a leader
.
.
**Function** *Initiate-CodeBlock*
    |  leader ← true;
    |  timer($t_p$) ← random integer $\in (0, t_p^{max}]$;
**Function** *Relay-CodeBlock*
    |  deactivate timer;
    |  leader ← false;
    |  record ping direction;
    |  calculate average of incoming pings;
    |  move towards incoming ping;

    **Algorithm 9:** Code block for primitive "Moving collectively"

---

Figure 14 shows a swarm aggregating towards a leader located at the far right end of the swarm,which steadily moves towards the right. While following the leader, the remaining swarm forms a line behind it, being lead away. This primitive can be viewed as "aggregation at a specific, moving agent". For choosing a leading agent, the primitive "leader election", which was earlier introduced, can be executed prior to this primitive.

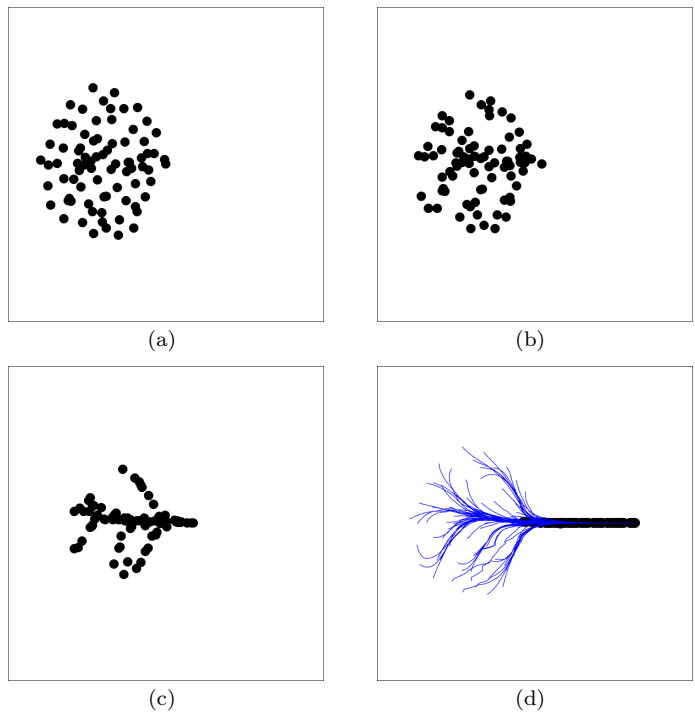

**Fig. 14** This figure shows a swarm being lead by a single agent towards the right. From the initial state in (a), the swarm aggregates towards the leading agent in (b) and (c). In (d) the swarm formed a line following the leader. Also shown in blue are the trajectories of each agent for the entire simulation. Times [$t_p^{max}$: (a) 1, (b) 4, (c) 12, (d) 40. Parameters: $N = 80$, $R_s = 3.34\,r$, $t_{ref} = 5\,s$, $t_p^{max} = 50$.

## 5.9 Locomotion: Gas expansion

Considering a swarm exploring its environment, it can maximize its covered surrounding area by physically expanding. The primitive "gas expansion" enables a swarm to uniformly expand. Each agent sets its internal counter $t_p$ to a random value between $t_p \in (0, t_p^{max}]$. As soon as the internal counter reaches $t_p = 0$ an agent sends a ping. Each agent moves away a small step from incoming pings. As soon as an agent receives no pings anymore, it does not move further away. See Algorithm 10 for pseudo code.

**Data:** Paradigm parameters
**Result:** Expanded swarm
.
.
**Function** *Initiate-CodeBlock*
  | timer($t_p$) ← random integer ∈ $(0, t_p^{max}]$;
**Function** *Relay-CodeBlock*
  | timer($t_p$) ← $t_p^{max}$;
  | record ping direction;
  | calculate average of incoming pings;
  | move away from incoming ping;

**Algorithm 10:** Code block for primitive "Gas expansion"

In Figure 15 (a) an initially densely packed swarm is shown, which then expands in Figure 15 (b) and (c) until it is fully expanded in Figure 15 (d). Agents can then reconnect with its swarm members by moving back, in the opposite direction of the previous step, or by integrating its entire trajectory and thus finding their way back until they again perceive signals. Depending on the communication abilities of the swarm, the perception range or sensitivity can be temporarily decreased during the expansion such that afterwards the agents will again be connected.

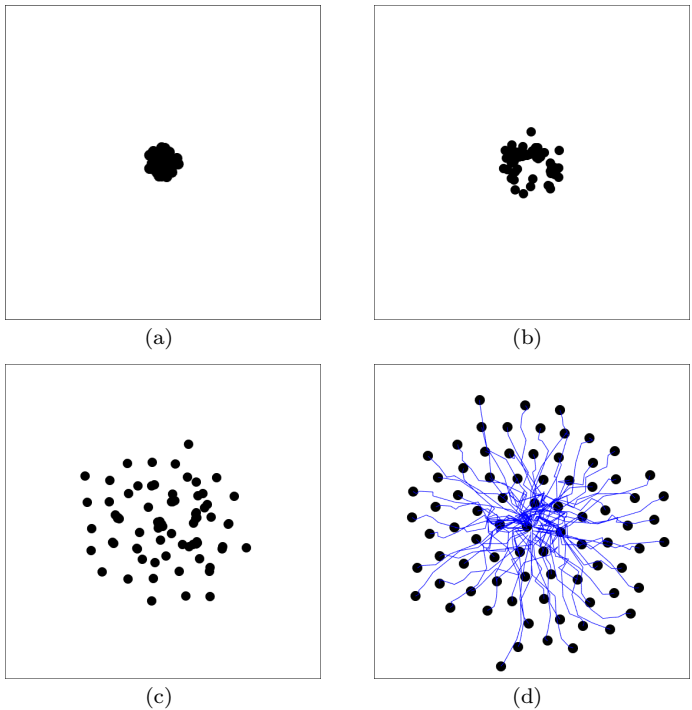

(a)          (b)

(c)          (d)

**Fig. 15** A swarm performing the primitive "gas expansion". The initial state of the swarm is shown in (a), where it is fully aggregated. In (b) and (c) it gradually expands. The final state is shown in (d) with the trajectories of each swarm as blue lines. Times [timesteps]: (a) 0, (b) 20, (c) 60, (d) 280. Parameters: $N = 80$, $R_s = 0.67\,r$, $t_{ref} = 5\,s$, $t_p^{max} = 50\,s$.

## 6 Combining primitives

All previously presented primitives are based on single-bit communication for demonstrating that even with minimum communication bandwidth complex behaviors can be produced. Using multi-bit signals the spectrum of possible primitives can be extended significantly. Examples and possibilities for multi-bit communication based systems and behaviors are presented in the discussion.

However, let alone by combining primitives not only more elaborate behaviors can be produced, by enabling a swarm to switch between a set of primitives it can operate autonomously. The most intuitive way of combining primitives is to execute primitives one after another in a sequential manner. This allows the design of complex tasks which can be executed by the swarm autonomously. This is schematically shown in Figure 16 (a).

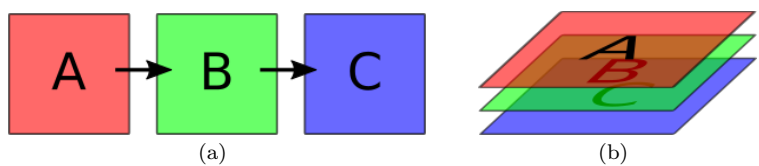

**Fig. 16** Schematic illustration of (a): sequential execution and (b): interleaved execution of three primitives A, B and C.

An example for sequential execution of primitives producing an autonomously acting swarm is a collective exploration procedure, shown in Figure 17. The following sequence of primitives is executed periodically: aggregation, leader election, moving collectively, gas expansion. In Figure 17(a)-(b) the swarm aggregates and then determines a leader in Figure 17(c). This leader will choose a random direction and lead the swarm to a new location, as shown in Figure 17(c) to (d). Then the entire swarm expands and explores the area and for example, collects data before again aggregating and restarting this procedure. Due to the limited abilities of the individual members of the swarm, they have no awareness of the collective state or if the execution of a primitve was completed. For this example, the execution times of all primitives were fixed or "hard-coded".

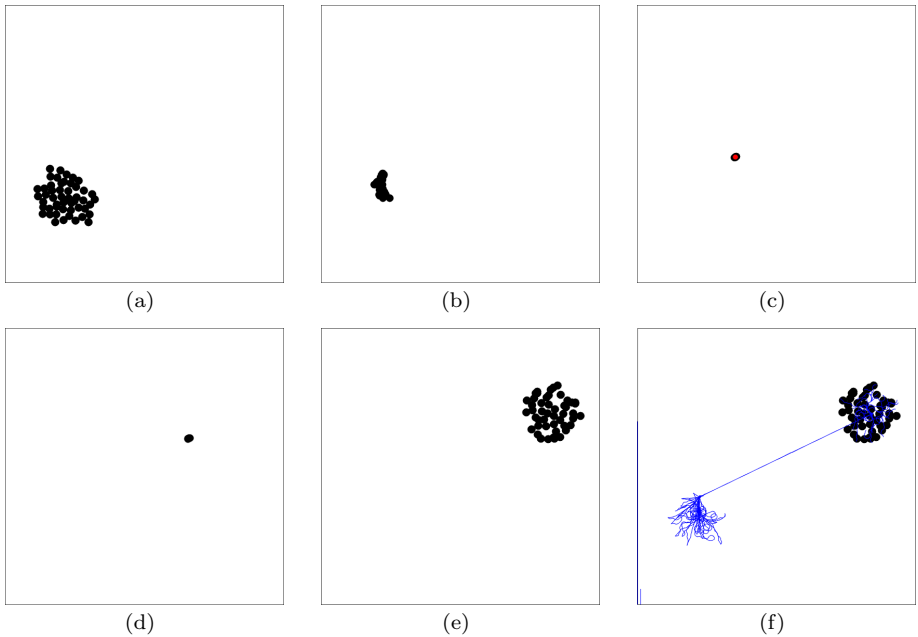

**Fig. 17** Consecutive execution of the primitives Aggregation, Leader election, Moving collectively and Gas expansion as example for an exploring routine of an autonomous swarm. The swarm prepares for changing its location and thus aggregates from (a) to (c). It then decides on a leading agent (marked in red) which then leads the swarm towards the top right of the system, a target area, shown in (c) and (d). The swarm expands again for exploring the new environment, shown in (e). In (f) the final state along with the trajectories of all agents over the entire simulation are shown. Parameters: $N = 50$, $t_{ref}$ and $t_p^{max}$ vary for each primitive.

Another approach to combining primitives is to execute several primitives in an interleaved manner. This allows the emergence of a larger variety of complex behaviors. This is schematically shown in Figure 16 (b). For executing several primitives in a quasi-simultaneous manner, the previously presented single bit communication can be extended. A simple option is to introduce several individual layers of single bit communication, one for each primitive. Alternatively, multi bit signals could be used, every signal encoding the primitive it is associated with.

The demonstrations shown here are only exemplary and many more combinations of primitives are possible. In case an observer controls a swarm, autonomy of the swarm is not necessary and full control through manual choice of primitives can be exerted.

## 7 Discussion

In Section 5, a set of primitives which can be utilized and combined as basic building blocks for a meta control scheme for a swarm is presented, covering the categories "internal organization", "swarm awareness" and "locomotion". Two exemplary realizations of combination of previously presented primitives for complex collective behaviors are presented in Section 6. Through this paper, it is demonstrated that WOSPP enables swarms consisting of agents with limited abilities to collectively perform a large variety of complex behaviors using "scroll wave" based communication. Due to its simple and flexible fundamental concept, this programming paradigm is applicable to a large spectrum of different types of swarms and environments while requiring minimal communication abilities. A cross section of possible abilities of WOSPP is presented in Section 5 where both primitives and basic analysis of their behaviors is shown. In this section WOSPP as a whole including its advantages, scope for further work and application in robotics is discussed.

In the introduction, several areas dealing with swarm control, such as amorphous computing were briefly examined. As pointed out, there are significant differences between amorphous computing and WOSPP one of which being that amorphous computing requires multi-bit communication. In amorphous computing, multiple gradients are propagated starting from a seed agent. The gradient is essentially a hop count which enables the internal positioning of agents with respect to the seed agent which in turn allows a group of robots to organize themselves globally. The paradigm presented in this paper functions even using single bit communication between the agents. Therefore, the extreme case i.e, the restriction to single-bit communication, is focused on in this paper. Though many systems do not have such constraint and could thus make use of complex signals being transmitted yielding even greater versatility. One such possibility is the use of so called "hop counts", encoding within the signal how often it was relayed, as done in [24, 1]. It allows for instance the limitation of the range up to how many agents or nodes a signal is relayed.

An analysis has been conducted on the resilience of "scroll wave" based communication[41] where its robustness against signal loss was examined. It was shown that due to redundancy in signal pathways, a system using slime mold based communication, as utilized in WOSPP, can compensate up to 70% individual probability of signal loss without significant decrease in performance. The ability of this basic behavior to cope with high amounts of signal loss endows WOSPP with resilient functioning when pings fail to be sent or received.

As opposed to approaches such as [22], decentralized control in WOSPP allows scalability limited primarily by the communication abilities relative to operational time scales. For the class of swarms presented in this paper the main constraint is constituted by the condition to choose the maximum internal cycle length $t_p^{max}$ significantly larger than the time for a ping wave to propagate from one end of the swarm to the other. It ensures that ping waves likely propagate through the entire system without colliding with other waves, thus enabling swarm-wide communication. By sufficient choice of parameters, the number of swarm members can be increased almost arbitrarily without loss of functionality.

However, considering a swarm encountering frequent collisions of ping waves, for instance while aggregating a swarm could ultimately split into sub-swarms, especially at sparsely connected regions in the swarm. Each group that splits out of the parent swarm retains the properties of the original swarm. Thus, in case of splitting, each sub-swarm immediately adapts to the new situation, being fully functional. Such properties of WOSPP can even be used to add to the richness in collective behavioral diversity. For example, swarms performing a search or exploration task can deliberately split into sub-swarms and proceed separately.

One of the prerequisites for a swarm to be able to implement WOSPP is directional communication similar to most animals in nature which exhibit swarm behavior. In this paper, most simulations conducted follow the assumption that agents have the ability to precise detect the direction of incoming pings. In practice, this requirement can be substantially loosened for the agents to have a lower angular resolution without significant loss of functionality. In other words, a rough perception of the direction of the incoming ping is sufficient for the basic functionality of WOSPP. Preliminary work was done but is not shown in this paper and will be subject of future work and further analysis on WOSPP. Only a selection of primitives is shown in this work, more ideas for primitives are for instance the incorporation of statistical regularities in very large swarms. The spectrum of collective behaviors can be varied and extended much further and applied a variety of systems, naturally depending on the requirements, tasks and creativity of the user. Consequently, the paradigm is applicable not only to swarms, but yields great potential in network system such as sensor networks, internet of things et cetera due to structural similarities in signal/information propagation throughout the system. In those examples presumably the main focus will be on "internal organization" and "swarm awareness" primitives.

Other future work connected to subCULTron[38], a project aiming to deploy a heterogeneous swarm of underwater robots to monitor environmental parameters in the lagoon of Venice. Within the framework of this project, individual primitives of WOSPP are already being used for swarm control. Robotic systems such as subCULTron, which employ a large number of individual agents in noisy environment aiming for autonomous operation, can benefit from WOSPP. In the future, a WOSPP language will be developed enabling users to combine primitives in a convenient manner and apply them as control scheme to a swarm of robots. Alternatively, a programming language for robotic swarms, called "buzz" [29] can be used for implementing the WOSPP. This will further facilitate the usage and application of WOSPP and the development of increasingly elaborate primitives, e.g. involving complex collective decision making, allowing in the future an easier designing of fully autonomous swarms with the ability to flexibly adapt to varying environmental conditions.

**Acknowledgements** This work was supported by EU-H2020 Project no. 640967, subCUL-Tron, funded by the European Unions Horizon 2020 research and innovation programme.